\documentclass[11pt]{article}
\usepackage{graphicx}
\newcommand{\half}{\frac{1}{2}}

\begin{document}
\title{``Back-of-the-envelope'' wind and altitude correction for 100 metre 
sprint times}

\author{J.\ R.\ Mureika\thanks{newt@palmtree.physics.utoronto.ca} 
\\
{\it Department of Physics} \\
{\it University of Toronto} \\
{\it Toronto, Ontario~~Canada~~M5S 1A7} }

\maketitle
\vskip .25 cm

\noindent
{\footnotesize
{\bf Abstract} \\
A simple algebraic expression is presented to correct men's and women's 100 
metre sprint times for ambient wind and altitude effects.  The simplified
formula is derived from a more complicated equation of motion used to
model the athlete's velocity as a function of time (the velocity curve).
This method predicts adjustments to 0-wind and 0-altitude equivalents, 
which are in excellent agreement 
to other estimates presented in the literature.  The expression 
is easily programmable on any computer, and could conveniently be used by 
coaches, meet directors, or the media to assess the performance of athletes and 
the quality of a race immediately following the event.
}
\pagebreak
\section{Introduction}

Although not officially recognized by the International Amateur Athletic
Federation (IAAF), correcting sprint race times for the effects of wind and 
altitude variation is a subject of increasing interest in the 
Track and Field community.   With the 
number of men's sub-9.90~s and women's sub-10.80~s clockings on the rise, 
correcting these marks to their sea-level, 0-wind equivalents is useful in 
determining the overall quality of the 
performances (at the time of competition).   A literature search reveals 
rather detailed experimental field studies of these effects 
\cite{dapena1,dapena2,davies,linth1,linth2}, as
well as several theoretical estimates based on 
mathematical and computational simulations \cite{ath,me100,pritch}.

Physically, linear drag forces (scaled to units of mass) are expressed as

\begin{equation}
F_d = \half C_d A \rho(H) ( v(t) - w )^2~,
\label{drag}
\end{equation}

where $C_d$ is the drag coefficient, $A$ the frontal cross-sectional area, 
$\rho(H)$ the 
atmospheric density at altitude $H$, $v(t)$ the sprinter's velocity, and $w$ 
the wind speed (co-linear to $v(t)$).   It follows that $F_d$ will be 
smaller for tail-winds ($w > 0$), and larger for head-winds ($w < 0$) 
at a fixed altitude.   Head-winds and tail-winds of equal magnitude will not 
provide equal-magnitude time corrections, due to the non-linear nature of the 
drag term. 

At 0 metres altitude with 0 wind, the base drag is  
$F_0 = 1/2 C_d A \rho_0 v(t)^2$, 
where $\rho_0 = 1.184~$g$\;$cm$^{-3}$ is the mean sea-level density of 
air at 25 degrees Celsius.  Since air density varies exponentially as a 
function of altitude, a convenient approximation can be written
as $\rho(H) = \rho_0 \exp(-0.000125 \cdot H)$ \cite{dapena1} for the 
range of elevations considered herein (less than 2300~m for the majority of 
competition venues).

The general consensus of most researchers in question is that for a 10.00~s 
race (average men's world-class sprint), a tail-wind of +2~ms$^{-1}$ 
will provide an advantage
of roughly 0.10~seconds ({\it i.e.} a faster time),
whose value will vary slightly depending on the altitude of the competition 
venue.  If the wind gauge reads in excess of +2~ms$^{-1}$, the 
performance is termed wind-assisted, and is not eligible for any potential
record status.  Conversely with no wind, an altitude of 1000~m will 
produce an advantage of 0.03~s, above which performances are officially deemed 
altitude-assisted.  Unlike wind-assisted marks, an altitude-assisted 
time can still count for a record.
At 2000~m, the advantage will be about 0.06~s over a 
sea-level run.  An 11.00~s time (average world-class women) will be boosted 
by about +0.12~s with a +2~ms$^{-1}$ tail-wind, and by 0.07~s (no wind) 
at 2000~m.  
As altitude increases, the magnitude of the wind effects will increase.  
Obviously, this is a reasonable explanation for the rash of World 
Records (WRs) experienced in the sprints and long jump at the 
1968 Olympics in Mexico City, which resides at an altitude of approximately 
2250~m.

\section{``Back-of-the-Envelope'' correction: Derivation}

A "back-of-the-envelope" (BOTE) calculation is a simplified reduction of a 
complex (physical) model, from which one can make reasonable predictions
with a minimal number of input parameters.  
An exact modeling of wind and altitude effects is a daunting task, since 
the mechanics involved are numerous and not easily representable by 
basic functions (see \cite{me100} for such a model).  
A historically-based method of such simulations is
via the velocity curve approach.  This is a method of studying a 
sprinter's performance, first introduced empirically by Hill \cite{hill} 
in the early 1900s, and further investigated 
by Keller \cite{keller} as an equation of motion of the form

\begin{equation}
\dot{v}(t) = F_p - v(t)\; \alpha^{-1}~,
\label{eq2}
\end{equation}
The term $F_p$ is a propulsive term, while $\alpha$ is a decay term
(representing internal physiological or biomechanical variables).
Note that again Equations~(\ref{eq2}) is scaled in units of the 
sprinter's mass $M$, so the interpretation of $F_d$ is force per unit mass 
(or, effectively, acceleration).  Unless otherwise specified, this notation 
is used for the remainder of the article.

This derivation roughly follows that of Reference~\cite{pritch}, however
the latter incorrectly estimates the numerical value of 
certain key variables, and omits the effects of altitude all together.
In fact, the author of \cite{pritch} suggests that altitude effects on
sprint times {\it cannot} be modeled by drag modification alone, which
is not necessarily a correct assertion (as will be shown).

Equation~(\ref{eq2}) may easily be altered to include drag effects by
the addition of $F_d$,
\begin{equation}
\dot{v}(t) = F_p - v(t)\; \alpha^{-1} - F_d~,
\label{eq3}
\end{equation}
and a time dependence $F_p \rightarrow F_p(t)$ may also be added
(see {\it e.g.} \cite{me100,mecjp,tibs} for such mechanisms).   
The BOTE expression presented herein, being simplistic by its namesake,
does not include these substitutions, and furthermore imposes an additional
simplification of time-independence, $v(t) \rightarrow v$, and
$\dot{v}(t) = 0$. 

To address the issue of wind and altitude correction, define as follows
$v(w,H)$ (velocity of the sprinter at altitude $H$ with wind $w$);
$v(0,0)$ (velocity with 0-wind, at sea-level); and $F_d(w,H)$ (effective
drag for wind $w$ and altitude $H$).  Subject to the constraint of
constant velocity, Equation~(\ref{eq3}) may be rewritten as

\begin{eqnarray}
v(w,H) &=& \alpha [F_p - F_d(w,H)]~,\nonumber \\
v(0,0) &=& \alpha [F_p - F_d(0,0)]~,
\end{eqnarray}
for each case described above.

Solving for $\alpha$ and equating the two expressions above yields

\begin{equation}
\frac{v(0,0)}{v(w,H)} = \frac{( 1 - F_d(0,0)/F_p )}{(1 - F_d(w,H)/F_p)} ~,
\label{eq4}
\end{equation}
Define the ratio $\delta = F_d(0,0)/F_p$ as the effort required to overcome 
drag in 0-wind conditions at sea level.  The numerical value of $\delta$
will be discussed shortly.  

Since velocity is constant ({\it i.e.} the average race velocity),
one can write $v(w,H) = 100 / t_{w,H}$,  where $t_{w,H}$ is the official 
time for the race under consideration, and rewrite Equation~(\ref{eq4}) as

\begin{equation}
\frac{t_{0,0}}{t_{w,H}} = \frac{( 1 - F_d(w,H)/F_p  )}{( 1 - \delta )}~,
\label{eq5}
\end{equation}

To simplify this expression further, note that the drag force for arbitrary $w$
and $H$ can be written as

\begin{eqnarray}
F_d(w,H)&=&\half C_d A \rho(H) v(w,H)^2 \left(1 - \frac{w}{v(w,H)} \right)^2 \nonumber \\
        &=&F_d(0,0) \left(\frac{v(w,H)}{v_{0,0}}\right)^2 \exp(-0.000125 \cdot H) 
\left( 1 - \frac{w \cdot t_{w,H}}{100}\right)^2 
\end{eqnarray}
So, replacing $(v(w,H)/v_0) = (t_{0,0} / t_{w,H})$, 

\begin{equation}
\frac{F_d(w,H)}{F_p} = \delta \left(\frac{t_{0,0}}{t_{w,H}}\right)^2 \exp(-0.000125\cdot H) 
\left( 1 - \frac{w \cdot t_{w,H}}{100} \right)^2~,
\end{equation}
and thus 

\begin{equation}
\frac{t_{0,0}}{t_{w,H}} = \frac{1}{(1-\delta)} \left[ 1 - \delta
\left(\frac{t_{0,0}}{t_{w,H}}\right)^2 exp(-0.000125\cdot H) 
\left( 1 - \frac{w \cdot t_{w,H}}{100)}\right)^2\right]~.
\label{eq6}
\end{equation}

Unfortunately, this is now a quadratic expression in $(t_{0,0}/t_{w,H})$, but
this problem is quickly resolved by making the following substitution.
Rewrite $(t_{0,0}/t_{w,H}) = 1 + \Delta t/t_{w,H}$, with $\Delta t =
t_{0,0} - t_{w,H}$.  Since $\Delta t$ will seldom be larger than
0.3~s for a $\sim$10~s race, it is reasonable to make the substitution

\begin{equation}
\left(\frac{t_{0,0}}{t_{w,H}}\right)^2 = \;
   \left(1 + \frac{\Delta t}{t_{w,H}}\right)^2 \;
  \simeq  1 + 2\; \frac{\Delta t}{t_{w,H}} \;
  =  2\; \frac{t_{0,0}}{t_{w,H}} - 1 
\label{subs1}
\end{equation}

The numerical value of $\delta$ is determined as

\begin{equation}
\delta = \frac{F_d(0,0)}{F_p} = \half \frac{C_d A \rho_0 v^2}{M\;F_p}
\end{equation}
(recall that the earlier definition of $F$ is scaled in units of inverse mass, 
hence the need for
$M$ in the denominator).  Pritchard \cite{pritch} initially found a value of 
$\delta \simeq 0.032$ (i.e. $3.2\%$ of a sprinter's effort is required to 
overcome drag), however this assumed an overestimated value of the drag 
coefficient $C_d = 1.0$, as well as the mean propulsive force 
$F_p = 12.1$~ms$^{-2}$.  Current research suggests a drag coefficient
of $C_d \in (0.5,0.6)$ \cite{me100,linth3}, as well as an average
$F_p \sim 7$~ms$^{-2}$ \cite{me100,mecjp,tibs}.

For a $9.90-10.00$~s race, the average velocity is between 
$v = 10 - 10.1$~m$\;$s$^{-1}$. Taking the drag area to be 
$C_d\cdot A = 0.23~$m$^2$
(consistent with the quoted $C_d$ values, and cross-sectional area
$A \in (0.4,0.5)$~m$^2$, for a sprinter of 
mass 75~kg, one finds $\delta \sim 0.027$.
  
Since $\delta$ is small, $1/(1-\delta) \simeq (1 + \delta)$,
and including the approximation of Equation~(\ref{subs1}),
Equation~(\ref{eq6}) may be rearranged as
\begin{eqnarray}
\frac{t_{0,0}}{t_{w,H}}&=& \left(\frac{1}{1-\delta}\right) 
\left[1 - \delta \exp(-0.000125\cdot H) \;
\left( 2\; \frac{t_{0,0}}{t_{w,H}}-1 \right) \;
\left(1 - \frac{w\cdot t_{w,H}}{100}\right)^2 \right]  \nonumber \\
& \simeq & 1 + \delta - \delta \exp(-0.000125\cdot H)\;
\left(1 - \frac{w\cdot t_{w,H}}{100}\right)^2 + o(\delta^2)~,
\end{eqnarray}
Thus, inserting the numerical value of $\delta$, one obtains the 
``back-of-the-envelope'' calculation
\begin{equation}
t_{0,0} \simeq  t_{w,H} [1.027 - 0.027 \exp(-0.000125\cdot H) (1 - w\cdot t_{w,H} / 100)^2 ] ~.
\label{eq8}
\end{equation}
For women, the input parameters $F_p$, $A$, $v^2$, and $M$ are smaller, 
so assuming values $v = (100/11) =  9.1~$ms$^{-1}$, $M = 65$~kg,
$F_p \sim 5$~ms$^{-2}$, and $A \sim 0.35~$m$^2$, $\delta$ remains essentially
unchanged.

Equation~(\ref{eq8})  
provides an excellent match to the predictions of Reference~\cite{me100}, 
as well as those of Dapena \cite{dapena2} and Linthorne \cite{linth2}.  
Thus, 100~metre sprint 
times may be corrected to their 0-wind, sea level equivalents by inputting 
only the official time, the wind gauge reading, and the altitude of the 
sporting venue.  Furthermore, Equation (8) is easily programmable in most 
scientific calculators and personal computers, and hence may be 
used track-side by coaches, officials and the media immediately following a 
race to gauge its overall ``quality''.

\section{Applications}

To demonstrate the utility of Equation~(\ref{eq8}),
Tables~\ref{legal},~\ref{windy},~\ref{altitude}, and \ref{headwind} 
present the corresponding corrections to the top five all-time 
men's and women's 100~m performances run with legal tail-winds, 
illegal winds ($w > +2.0$~ms$^{-1}$), altitude effects ($H > 1000$~m), 
and extreme head-winds ($w < -1~$ms$^{-1}$).

The current 100~m World Record (as of June 2000) of 9.79~s by Maurice
Greene was run at low altitude with virtually no wind, and adjusts to
9.80~s.  Note that (Table~\ref{legal}) the 9.86~s performances of Trinidadian
Ato Boldon and Namibia's Frank Fredericks were both run into head-winds
of equal magnitude, but the altitude difference allows for a 0.02~s 
differential in corrected times.  The former World Record of 9.84~s by 
Canada's Donovan Bailey corrects to a 9.88~s.  
It is also interesting to note how exceptional performances can be hampered by
strong head-winds (Table~\ref{headwind}). 

The 10.49~s WR of the late Florence Griffith-Joyner is included in both 
Table~\ref{legal} and Table~\ref{windy}, to demonstrate the common belief that 
this mark was strongly wind-aided (despite the fact that the official wind
gauge reading was +0.0~ms$^{-1}$, there is strong circumstantial evidence to
suggest that the equipment malfunctioned).  Griffith-Joyner's legal personal 
records 
(PRs) correct to about 10.68~s, while her +2.1~ms$^{-1}$ wind-aided 10.54~s 
(Seoul, 1988)  corrects to 10.66~s.  Thus, the actual WR mark should probably 
be 10.60-10.65~s (or a 0-wind, 0-altitude equivalent of about 10.66-10.68~s).  
American Marion Jones' current adjusted PR is 10.69~s, effectively on par
with Griffith-Joyner's best marks.

From a historical perspective, it is interesting to note that Calvin 
Smith's 10.04~s ($-$2.2~ms$^{-1}$) in 1983 would have converted to a 0-wind 
9.93~s at sea level.  Smith's actual WR of 9.93~s (+1.2~ms$^{-1}$) was run at 
altitude (Colorado Springs, USA; 1850~m), correcting to only 10.03~s.
The former WRs of Canada's Ben Johnson, 9.79~s (+1.1~ms$^{-1}$) in 
Seoul, SKR, and 9.83~s (+1.0~ms$^{-1}$) in Rome, ITA, would correct to
9.85~s and 9.88~s, respectively.  Unfortunately, these marks were stricken
as a result of performance-enhancing drug infractions.

\section{Conclusions}
The presented ``back-of-the-envelope'' calculation is simple to use, and
is applicable to both men and women's performances.  An on-line JavaScript
version is available at the author's website, currently

\begin{center}
{\tt http://palmtree.physics.utoronto.ca/$\sim$newt/track/wind/}
\end{center}

It is hoped
that its use may be eventually adopted by the IAAF and/or other governing 
bodies of Athletics as a relative gauge of performance quality under 
differing competition conditions.

\vskip .3cm
\noindent{\Large {\bf Acknowledgements}} \\
I thank Jesus Dapena for helpful discussions, and for providing the data of 
Reference~\cite{dapena2} prior to its publication.  
This work was supported in part by a 
Walter C.\ Sumner Memorial Fellowship, as well as a grant from the National 
Sciences and Engineering Research Council.

\pagebreak
\begin{table}[h]
\begin{center}
{\begin{tabular}{l r l l r}\hline
Athlete &$t_{w,H}$ ($w$) & Venue (Altitude) & Date & $t_{0,0}$ \\ \hline
Maurice Greene USA & 9.79 (+0.1) & Athens, GRE (110) &99/06/16& 9.80 \\
Bruny Surin CAN & 9.84 (+0.2) & Athens, GRE  & 99/08/22&9.85 \\
Donovan Bailey CAN & 9.84 (+0.7) & Atlanta, USA (315) &96/07/27& 9.88 \\
Leroy Burrell USA & 9.85 (+1.2) & Lausanne, SWI (600) &94/07/06& 9.92 \\
Ato Boldon TRI & 9.86 ($-$0.4) & Athens, GRE  & 98/06/17& 9.84 \\
Frank Fredericks NAM & 9.86 ($-$0.4) & Lausanne, SWI & 96/07/03& 9.86 \\ \hline
Florence Griffith-Joyner USA&10.49 (+0.0)&Indianapolis, USA (220)&88/07/16&10.50 \\
&10.61(+1.2)&Indianapolis, USA &88/07/17&10.68\\
Marion Jones USA &10.72 (+0.0)&Monaco  (10)&98/08/08&10.72\\
Christine Aaron   FRA&10.73 (+2.0)&Budapest, HUN (150)&98/08/19&10.84 \\
Merlene Ottey  JAM & 10.74 (+1.3)& Milano, ITA (121) & 96/09/07 & 10.82 \\
Evelyn Ashford USA & 10.76 (+1.7)& Zurich, SWI (410) & 84/08/22 & 10.87 \\ \hline
\end{tabular}}
\end{center}
\caption{
Men's and Women's top 5 all-time legal 100~m performances (best  
per athlete).  Times measured in seconds (s), and wind-speeds in ms$^{-1}$.
Altitude is assumed to be correct to within $\pm$20~m.
}
\label{legal}
\end{table}

\begin{table}[h]
\begin{center}
{\begin{tabular}{l r l l r}\hline
Athlete &$t_{w,H}$ (w) & Venue (Altitude) & Date & $t_{0,0}$ \\ \hline
Obadele Thompson  BAR &9.69 (+5.7) & El Paso, USA (1300) & 96/04/13 & 9.91 \\
Carl Lewis USA  & 9.78 (+5.2)&Indianapolis, USA & 88/07/16 &9.98 \\
Andre Cason  USA &9.79 (+4.5)& Eugene, USA& 93/06/16& 9.97 \\
Maurice Greene USA&9.79 (+2.9)&Eugene, USA&98/05/31&9.92 \\
Leonard Scott  USA&9.83 (+7.1)&Knoxville, USA (270)&99/04/09&10.07 \\ \hline
Griffith-Joyner USA&10.49 (+5.5)&Indianapolis, USA &88/07/16&10.72 \\
 & 10.54(+3.0) & Seoul, SKR (85)& 88/09/25 & 10.69  \\
Marion Jones  USA &10.75 (+4.1)&New Orleans, USA (10) &98/06/19 &10.95 \\
Gail Devers    USA&10.77 (+2.3)&San Jose, USA (10)&94/05/28&10.90 \\
Ekaterini Thanou   GRE&10.77 (+2.3)&Rethymno, GRE (20)&99/05/29&10.90 \\
Evelyn Ashford   USA&10.78 (+3.1)&Modesto, USA (25)&84/05/12&10.94 \\ \hline
\end{tabular}}
\end{center}
\caption{
Top 5 all-time wind-assisted marks ($w > +2.0$~ms$^{-1}$).
}
\label{windy}
\end{table}

\begin{table}[h]
\begin{center}
{\begin{tabular}{l r l l r}\hline
Athlete &$t_{w,H}$ (w) & Venue (Altitude) & Date & $t_{0,0}$ \\ \hline
Obadele Thompson BAR&9.87 ($-$0.2)&Johannesburg, RSA  (1750)&98/09/11&9.91 \\
Seun Ogunkoya NGR &9.92 ($-$0.2) &Johannesburg, RSA & 98-09-11 &9.96 \\ 
Calvin Smith USA & 9.93 (+1.4) & Colorado Springs, USA (1853)  & 83/07/03 & 10.04 \\
Jim Hines USA & 9.95 (+0.3) & Mexico City, MEX (2250) & 68/10/14 & 10.03 \\
Olapade Adeniken NGR & 9.95 (+1.9) & El Paso, USA (1300) & 94-04-16 & 10.07 \\ \hline
Marion Jones    USA&10.65 (+1.1)&Johannesburg, RSA &98/09/12&10.76 \\
Dawn Sowell  USA & 10.78 (+1.0)& Provo, USA (1380)& 89/06/03 & 10.81 \\
Evelyn Ashford  USA &10.79 (+0.6) & Colorado Springs, USA & 83/07/03 & 10.88 \\
Diane Williams  USA& 10.94 (+0.6) & Colorado Springs, USA & 83/07/03 & 11.03 \\
Chandra Sturrup BAH&10.97 (+1.1)& Johannesburg, RSA & 98/09/12 & 11.08 \\ \hline
\end{tabular}}
\end{center}
\caption{
Top 5 all-time altitude-assisted marks ($H > 1000$~m).
}
\label{altitude}
\end{table}

\begin{table}[h]
\begin{center}
{\begin{tabular}{l r l l r}\hline
Athlete &$t_{w,H}$ (w) & Venue (Altitude) & Date & $t_{0,0}$ \\ \hline
Maurice Greene   USA&9.96 ($-$1.0)&Uniondale, USA  (30)&98/07/21&9.92\\
Leroy Burrell    USA&9.97 ($-$1.3)&Barcelona, ESP (95)&92/08/01&9.90\\
Linford Christie   USA&10.00 ($-$1.3)&Barcelona, ESP &92/08/01&9.93\\
Ato Boldon   TRI&10.00 ($-$1.0)&Uniondale, USA &98/07/21&9.96 \\
Donovan Bailey   CAN&10.03 ($-$2.1)&Abbotsford, CAN (40)&97/07/19&9.91\\ \hline
Irena Privalova  USR&10.84 ($-$1.0)&Barcelona, ESP&92/08/11&10.77 \\
Gwen Torrence  USA&10.86 ($-$1.0)&Barcelona, ESP&92/08/01&10.79 \\
Merlene Ottey    JAM&10.88 ($-$1.0)&Barcelona, ESP&92/08/01&10.81 \\
Evelyn Ashford  USA&10.96 ($-$1.4)&Knoxville, USA&82/06/19&10.88 \\
Jones    USA&10.97 ($-$1.1)&Indianapolis, USA&97/06/13&10.90 \\ \hline
\end{tabular}}
\end{center}
\caption{
Top 5 all-time marks with $w \leq -1.0$~ms$^{-1}$.
}
\label{headwind}
\end{table}

\end{document}